\pdfoutput=1
\documentclass[%
 reprint,
superscriptaddress,
 amsmath,
 amssymb,
 aps,
 showkeys,
 pra
]{revtex4-2}

\usepackage{graphicx}
\usepackage{bm}
\usepackage{amsmath}

\usepackage{hyperref}

\usepackage{gensymb}
\usepackage{soul}
\usepackage[version=4]{mhchem}
\usepackage{xcolor}

\renewcommand{\vec}{\bm}

\newcommand{\avg}[1]{\left\langle#1\right\rangle}
\newcommand{\E}[2][]{\mathbb{E}\left[#2\right]_{#1}}
\newcommand{\V}[1]{\text{Var}\left(#1 \right) }
\newcommand{\C}[1]{\text{Cov}\left(#1 \right) }

\def\tv{\tau_{v}}
\def\tmopt{\tau_{\textrm m}^{\rm opt}}
\def\tm{\tau_{\textrm m}}
\def\tr{\tau_{\textrm r}}
\def\tc{\tau_{\textrm c}}

\def\xT{X_{\textrm T}}
\def\RT{R_{\rm T}}

\def\mN{\overline{N}}
\def\Neff{\overline{N}_{\textrm{eff}}}
\def\NI{\overline{N}_{\textrm I}}
\def\NIppn{\bar N_{\textrm I}^{\textrm{PPN}}}
\def\fI{f_{\textrm I}}

\def\snrinv{\text{SNR}^{-1}}

\def\Nr{N_\mathrm{r}}
\def\vt{v_\tau}

\def\st{s_\tau}
\def\sp{\vec{s}}
\def\hst{\hat s_\tau}
\def\sst{\sigma_{s_\tau}^2}
\def\ph{\hat{p}_{\tau_r}}

\def\dg{\tilde{g}}

\def\dgs{\tilde{g}^2}

\def\ep{\sigma^2_{\hat p_{\tau_{\textrm{r}}}}}

\def\RI{R_{\textrm{I}}}

\def\KDI{{K_{\rm D}^{\rm I}}}
\def\KDA{{K_{\rm D}^{\rm A}}}

\newcommand{\fref}[1]{\textcolor{blue}{Fig.~\ref{fig:#1}}}

\newcommand{\flabel}[1]{\label{fig:#1}}
\newcommand{\eref}[1]{Eq.~\ref{eqn:#1}}
\newcommand{\erefstart}[1]{Equation~\ref{eqn:#1}}

\newcommand{\erefstwo}[2]{Eqs.~\ref{eqn:#1}~and~\ref{eqn:#2}}
\newcommand{\erefsthree}[3]{Eqs.~\ref{eqn:#1}, ~\ref{eqn:#2}~and~\ref{eqn:#3}}
\newcommand{\erefsrange}[2]{Eqs.~\ref{eqn:#1}-\ref{eqn:#2}}
\newcommand{\elabel}[1]{\label{eqn:#1}}

\begin{document}

\title{Predicting concentration changes via discrete sampling}

\author{Age J. Tjalma}
\affiliation{AMOLF, Science Park 104, 1098 XG Amsterdam, The Netherlands}
\author{Pieter Rein ten Wolde}
\email{tenwolde@amolf.nl}
\affiliation{AMOLF, Science Park 104, 1098 XG Amsterdam, The Netherlands}

\date{\today}

\begin{abstract}
To successfully navigate chemical gradients, microorganisms need to predict how the ligand concentration changes in space. Due to their limited size, they do not take a spatial derivative over their body length but rather a temporal derivative, comparing the current signal with that in the recent past, over the so-called adaptation time. This strategy is pervasive in biology, but it remains unclear what determines the accuracy of such measurements. Using a generalized version of the previously established sampling framework, we investigate how resource limitations and the statistics of the input signal set the optimal design of a well-characterized network that measures temporal concentration changes: the {\it Escherichia coli} chemotaxis network. Our results show how an optimal adaptation time arises from the trade-off between the sampling error, caused by the stochastic nature of the network, and the dynamical error, caused by uninformative fluctuations in the input. A larger resource availability reduces the sampling error, which allows for a smaller adaptation time, thereby simultaneously decreasing the dynamical error. Similarly, we find that the optimal adaptation time scales inversely with the gradient steepness, because steeper gradients lift the signal above the noise and reduce the sampling error. These findings shed light on the principles that govern the optimal design of the {\it E. coli} chemotaxis network specifically, and any system measuring temporal changes more broadly.
\end{abstract}

\maketitle
\section{Introduction}
 Organisms ranging from bacteria to mammals have learned to navigate their environment in order to find food and avoid threats. Successful navigation requires the organism to predict the spatial structure of its surroundings, which necessitates measuring and storing relevant environmental properties. Therefore, how accurately these signals are sensed can fundamentally limit the success of navigation \cite{mattingly_escherichia_2021}. This in turn raises the question how accurately such signals can be transduced.

Microorganisms that navigate chemical gradients need to determine the correct direction to move in, which entails predicting the change in concentration that they will encounter, rather than its value. Because these organisms are typically small relative to the gradient length, the measurement error is large compared to the concentration difference over their body length \cite{berg_physics_1977}. Therefore, they cannot directly measure the gradient. Instead, these micro-organisms only have access to the local concentration. Yet, they can also store past concentrations. How these cells should integrate the current and past information to predict the concentration change remains however unclear. In principle, cells can combine the concentration value with its derivative to predict the concentration change, and the optimal strategy for combining this information depends on the statistics of the environment. If the range of background concentrations is large compared to the typical concentration change over the signal correlation time as set by the organism's own motion, then the optimal system for predicting the concentration change is one that exhibits perfect adaptation \cite{tjalma2023trade}. It means that the organism bases its prediction on the concentration change only.

Interestingly, various organisms have indeed been shown to employ this strategy. A canonical example is the bacterial chemotaxis system, which is widely conserved across species \cite{macnab1972gradient,segall_temporal_1986, lux2004chemotaxis,baker2006systems,rao2008three}. But also eukaryotic sperm cells measure temporal changes when navigating towards an egg \cite{kaupp2003signal, friedrich2007chemotaxis, abdelgalil2022sea}, and even the multicellular nematode {\it Caenorhabditis elegans} depends on temporal derivatives in a range of taxis behaviors \cite{lockery2011computational}.

Even though measuring temporal changes appears to be a common and important function, it is not clear what sets the accuracy of such measurements. The fundamental information processing devices that allow living cells to measure concentration changes are biochemical signaling networks. Like any device, the accuracy of such networks is limited by the physical resources required to build and operate them, such as energy, components, and time. Here, we investigate how these resources limit the accuracy with which cells can predict changes in the encountered concentration during navigation. Specifically, we ask what determines the optimal design of the signaling network under limited resource availability?

To measure a temporal change, cells subtract from the most recent signal the signal further back into the past. The latter is performed via the adaptation system. Crucially, to yield a response of non-zero amplitude, which is necessary to lift the signal above the inevitable biochemical noise, the system cannot adapt instantly; it therefore cannot take an instantaneous derivative. On the other hand, the adaptation time should not be too long, because then the temporal derivative is taken over a larger window stretching further back into the past, which is less informative about the current or future derivative that the cell needs to predict. We thus expect that there exists an optimal adaptation time that arises from this trade-off between a derivative that is most recent and one that is most reliable \cite{tjalma2023trade}. However, what precisely controls the optimal adaptation time, and how this depends on the statics of the input and the available resources such as receptor and readout copies, remains unknown.

An intuitive perspective that is ideally suited to answer these questions is the previously established sampling framework \cite{govern_optimal_2014, malaguti_theory_2021,malaguti_receptor_2022}. This framework views the signaling network downstream of the receptor as a device that discretely samples the state of the receptor. From this starting point, it enables identification of the different contributions that comprise the full sensing error: the sampling error, caused by fluctuations in the number of samples, the binary nature of the receptor state, and receptor-level noise; and the dynamical error, resulting from uninformative fluctuations in the input. While previous work has used the sampling framework to investigate sensing the current signal, we here generalize and extend it to include the prediction of signal properties a specified time into the future. We then apply this generalized sampling framework to the {\it Escherichia coli} chemotaxis network: a well-characterized example of a network which measures temporal changes. We model the input signal after the experimentally measured input for {\it E.coli} chemotaxis in shallow gradients \cite{mattingly_escherichia_2021}.

Our results distinctly show how an optimum for the adaptation time arises from its opposing effects on the sampling error and the dynamical error. While the former decreases with the adaptation time, the latter increases with it. Given the adaptation time, a larger number of receptor and readout molecules reduces the sampling error, shifting the balance between the sampling and dynamical error. Therefore, increasing the resource availability reduces the optimal adaptation time. Similarly, we find that the optimal adaptation time scales inversely with the steepness of the chemical gradient in which the organism navigates. The reason is that in a steeper gradient, the signal is more easily distinguished from the noise under the same resource availability. This again means that the sampling error decreases relative to the dynamical error, reducing the optimal adaptation time to decrease the latter. Finally, if the dynamics of the concentration change are Markovian, the optimal adaptation time is independent of the prediction interval. These findings likely extend well beyond {\it E. coli}, and have implications for the optimal design of any system that measures temporal changes, be it natural or man-made.

\begin{figure*}[t]
	\centering
	\includegraphics[width=\linewidth]{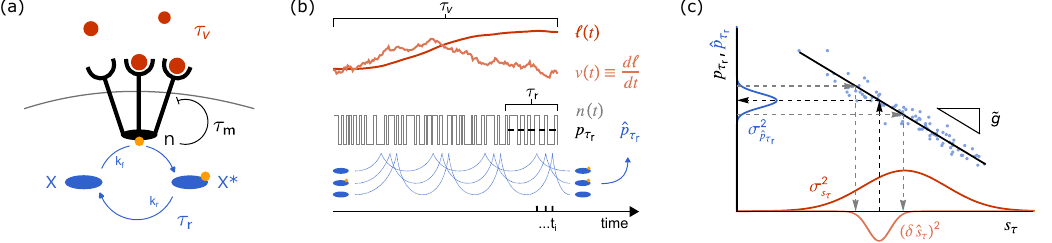}
	\caption{A push-pull motif samples the binary state of the chemotaxis receptor cluster. (a) Ligand binding affects the probability of a chemotaxis receptor cluster to reside in its active or inactive conformation. This binary cluster state $n$ controls the methylation dynamics of its constituent receptors, leading to negative feedback on the adaptation timescale $\tm$. The cluster state is sampled by the readout molecules $X$ on the response timescale $\tr$. (b) We consider an input signal defined by its concentration $\ell(t)$ and concentration derivative $v(t)$, with correlation time $\tv$ (\erefsthree{hol}{hov}{vcor1}). The instantaneous cluster activity $n\in \{0,1\}$ switches fast relative to the input correlation time, response time $\tr$, and adaptation time $\tm$. Due to the negative feedback, the mean cluster activity reflects the change in concentration over the past adaptation time $\tm$. The network makes an estimate $\ph$ of the cluster activity over the past response time $\tr$ by discretely sampling the instantaneous cluster state via the push-pull motif (panel (a)); the estimate $\ph=x^*/\mN$ is given by the current number of active readout molecules $x^*$, reflecting the number of samples of active receptor clusters during the past integration time $\tr$, over the mean number of samples $\mN$ during this time $\tr$ (\eref{phat}).(c) For linear Gaussian systems the future signal $\st$ maps onto a current mean cluster activity over the response time $p_{\tr}$ via the dynamic input-output relation of \eref{io}. The variance in the estimate $\ph$ given a signal value $\st$ is the prediction error $\ep$. Mapping the prediction error back onto the signal gives the network's error in the signal estimate $(\delta \hst)^2$. The ratio between the total variance in the signal $\sst$ and the error in the signal estimate $(\delta \hst)^2$ is the signal-to-noise ratio (\eref{SNRdef}). }
	\flabel{intro}
\end{figure*}

\section{Results}
\subsection{Theory: sampling framework}
In general, the function of a biochemical signaling network is to estimate the value of a signal of interest, which typically varies in time. Sensing entails estimating the value of the signal at the current time $t_0$, while predicting the future state of the environment implies estimating the value a time $\tau$ into the future. To extend the sampling framework to be applicable to prediction as well as sensing we define the signal of interest as $\st \equiv s(t_0+\tau)$ with $\tau\geq 0$. In this work we consider a time-varying input signal described by stationary Gaussian statistics (see section \ref{sec:sig}).

In biochemical signaling networks, the activity state of receptor proteins is altered by ligand molecules that bind them. In turn, downstream readout proteins stochastically sample the receptor state $n \in \{0,1\}$. From these samples the signal of interest must then be inferred. A canonical motif that samples the activity state of upstream receptor proteins is the push-pull network \cite{goldbeter_amplified_1981}. In this network a sample of the receptor state is stored in the chemical modification state of a readout protein, which decorrelates from the receptor state over the response time $\tr$ [\fref{intro}(a)]. 

To estimate the signal value $\st$ a time $\tau$ into the future, the cell integrates the receptor activity over a time $\tr$, leading to an estimate $\ph$ of the average receptor activity $p_{\tr}$ over the integration time $\tr$ [\fref{intro}(b)]. However, during this past time $\tr$, the input signal varies over its own timescale $\tv$, which leads to changes in the receptor activity on this timescale as well \cite{malaguti_theory_2021, malaguti_receptor_2022}. On top of variation on the timescale of the input dynamics, the receptor activity fluctuates on the timescale of ligand binding and unbinding, and on the timescale of the adaptation mechanism $\tm$. In the linear regime, the dynamic input-output relation between the average receptor activity $p_{\tr}$ and the signal of interest $\st$ is given by
\begin{align}
	p_{\tr}(\st)\equiv \E[t_i]{\avg{n(t_i)|\st}}= p+\dg \st, \elabel{io}
\end{align}
where the angle brackets denote an ensemble average over all receptors, $\E[t_i]{\dots}$ is an average over all sampling times $t_i$, which are exponentially distributed over the integration time $\tr$ (\eref{tidist}), and $p\equiv\E[t_i]{\avg{n(t_i)}}$ is the average receptor activity over all signal values. The dynamic input-output relation thus gives the average receptor activity $p_{\tr}$ over the response time $\tr$ given that the future signal is $\st$; $p_{\tr}$ is thus an average over all sources of noise, arising from receptor-ligand binding and receptor methylation, readout activation, and fluctuations in the past input that are not informative because they map onto the same future signal $\st$ (see \fref{error}). The slope of the mapping between $\st$ and $p_{\tr}$ is the dynamic gain $\dg$ [\fref{intro}(c)] \cite{tostevin_mutual_2010}.

The accuracy of any signaling device can be quantified using the signal-to-noise ratio (SNR), which is a measure for the number of distinct signal values the system can distinguish. For systems with Gaussian statistics, as studied here, the SNR is given by the ratio of the signal variance $\sst$ over the error in the cell's estimate of the signal $(\delta \hst)^2\equiv\E[\st]{\V{\hst | \st}}$, i.e. the variance of the cell's signal estimate $\hst$ under a fixed signal $\st$, averaged over all $\st$:
\begin{align}
	\text{SNR} \equiv \frac{\sst}{(\delta \hat{s}_\tau)^2} = \frac{\dg^2\sst}{\ep}. \elabel{SNRdef}
\end{align} 
The cell estimates the signal $\st$ from the average receptor activity over the integration time, $p_{\tr}$, via the dynamic input-output relation, see \eref{io} and \fref{intro}(c). Using the rules of error propagation, the error in the signal estimate is thus given by
\begin{align}
	(\delta \hat{s}_\tau)^2=\ep/\dgs, \elabel{errprop}
\end{align}
where the error in the estimate of the receptor activity $\ph$ over the integration time $\tr$ is defined as
\begin{align}
	\ep \equiv \E[\st]{\V{\ph | \st}} \elabel{errdef}
\end{align}
The signal to noise ratio of \eref{SNRdef} also specifies the Gaussian mutual information between the signal and the network output \cite{bialek_biophysics_2012}.

To quantify the error in the cell's estimate of the receptor activity (\eref{errdef}), we have to consider how the cell makes this estimate. As a model system to investigate networks that measure changes in the input we use the {\it E. coli} chemotaxis network. In this network, the activity of a receptor cluster reflects the change in signal concentration over the past adaptation time $\tm$ (see section \ref{sec:chem} for details). Downstream of the cluster, its activity state is sampled via a push-pull motif [\fref{intro}(a)] \cite{goldbeter_amplified_1981}. The cell's estimate of the fraction of active clusters is given by (also see \cite{govern_optimal_2014})
\begin{align}
	\ph = \frac{1}{\mN} \sum^N_{i=1} n_i(t_i) = \frac{x^*}{\mN}, \elabel{phat}
\end{align}
where $n_i(t_i) \in \{0,1\}$ is the outcome of sample $i$ at sampling time $t_i$, which is set by the binary activity state of the receptor cluster that was sampled at time $t_i$ [\fref{intro}(b)]. The physical readout of the network is the number of active readout molecules $x^*=\sum_{i=1}^N n_i(t_i)$, which have been phosphorylated by an active receptor cluster. Since readout phosphorylation is driven by ATP hydrolysis, we consider the sampling process in the irreversible limit.

The number of samples $N$ is set by the rate of sampling $r$ and the timescale over which samples remain correlated with the receptor state, which is set by the integration, or response time $\tr$. In the push-pull motif the sampling rate is set by the forward rate constant $k_f$, the number of receptor clusters $\RT$, and the number of available readout molecules $X$: $r=k_f x \RT$ [\fref{intro}(a)]. We assume that $N$ is Poisson distributed with mean $\mN  = \bar r \tr$. This mean number of samples can be expressed in terms of the steady state fraction of phosphorylated readouts $f=k_f p \RT \tr$ and the total number of readouts $\xT$ \cite{malaguti_receptor_2022},
\begin{align}
	\mN=f(1-f)\xT/p. \elabel{mn}
\end{align} 
The steady state flux of readout molecules is given by $\bar r p = f(1-f)\xT/\tr$.

Using the definition of the cell's estimate of the receptor activity (\eref{phat}) the error in this estimate (\eref{errdef}) can be decomposed into independent parts in a very general manner. We set out this decomposition in the section that follows. After the decomposition of the error we describe the dynamics and statistics of the input of the chemotaxis network (section \ref{sec:sig}). Subsequently we introduce the chemotaxis network in more detail, and compute the dynamic gain $\dg$ (see \eref{io}), and the different contributions to the error in terms of the parameters of the system (section \ref{sec:chem}). We then compute the full expression for the SNR and investigate its behavior as a function of the prediction interval, the resource availability, and the adaptation time (sections \ref{sec:snr}-F). Finally, we compare the predictions of our theory to available experimental data on the {\it E. coli} chemotaxis network (section \ref{sec:exp}).

\subsubsection*{The error in the estimate of the receptor activity}
We can derive a general expression for the prediction error $\ep$, which shows how the complete error decomposes into independent parts. We start from the definition of the error (\eref{errdef}), which we rewrite using the law of total variance
\begin{align}
	&\ep =\V{\ph} - \V{\E{\ph|\st}}, \nonumber \\
	&= \V{\E{\ph|N}} + \E{\V{\ph|N}} - \V{\E{\ph|\st}}, \elabel{errgen}
\end{align}
where in the first line we use that the total variance in the estimate of the activity $\V{\ph}$, is the sum of the variance in the mean of $\ph$ given $\st$, $\V{\E{\ph|\st}}$, and the mean of the variance in $\ph$ conditional on $\st$, $\E{\V{\ph|\st}}$, which is the error $\ep$ (\eref{errdef}). Indeed, the error in the estimate is its total variance minus the part which is informative about the signal of interest $\st$. Subsequently, in the second line, we split the total variance in the estimate $\ph$ into a part that arises from fluctuations in the number of samples $N$, the first RHS term, and the mean variance in $\ph$ when $N$ is fixed, the second RHS term.

In appendix \ref{app:prederr} we show how each term of \eref{errgen} can be simplified further using the definition of the cell's estimate $\ph$ (\eref{phat}). The first term, the error caused by fluctuations in the number of samples, is given by
\begin{align}
	 \V{\E{\ph|N}} = \frac{p^2}{\mN}, \elabel{varbyN}
\end{align}
with the average cluster activity $p\equiv\E[t_i]{\avg{n(t_i)}}$. As shown in previous work, this error would be zero if the sampled cluster functions bidirectionally, i.e. if inactive clusters would dephopshorylate readout molecules \cite{govern_optimal_2014}; in contrast, in the chemotaxis network deactivation is not driven by inactive receptor clusters but rather by an enzyme (CheZ) independent of the receptor state, and then this term is non-zero. The fluctuations under a fixed number of samples, the second RHS term of \eref{errgen}, can be decomposed further into three parts:
\begin{align}
 	\E{\V{\ph|N}} =& \frac{p(1-p)}{\mN} \nonumber \\
 	&+ \E[t_i,t_j,\sp]{\C{n_i(t_i),n_j(t_j)|\sp}} \nonumber \\
 	&+ \V{\E[t_i]{\avg{n(t_i)|\sp}}}, \elabel{varnoN}
\end{align} 
where the first part reflects the instantaneous variance of each sampled cluster, the second part is the cluster covariance under a fixed past signal trajectory $\sp\equiv \{s(t)\}_{t\leq t_0}$, and the third part quantifies the effect of the signal history $\sp$ on the activity of the cluster. Finally, the variance that is informative of the future signal value, i.e. the third RHS term of \eref{errgen}, is given by
\begin{align}
	\V{\E{\ph|\st}} = \V{\E[t_i]{\avg{n(t_i)|\st}}} = \dgs \sst, \elabel{varst}
\end{align}
which follows directly from the dynamic input output relation in \eref{io}. Substituting \erefsthree{varbyN}{varnoN}{varst} into \eref{errgen} yields the full prediction error  
\begin{widetext}
	\begin{align}
		\ep = \underbrace{\frac{p^2}{\mN} + \frac{p(1-p)}{\mN} + \E[t_i,t_j,\sp]{\C{n_i(t_i),n_j(t_j)|\sp}}}_\text{sampling error} + \underbrace{\V{\E[t_i]{\avg{n(t_i)|\sp}}}-\dgs \sst}_\text{dynamical error}. \elabel{prederr}
	\end{align}
\end{widetext}
The first three terms together make up the sampling error. This error arises due to the stochastic nature of the sampling process downstream of the receptor, receptor-ligand binding and unbinding, and the adaptation mechanism. In this work we integrate out ligand binding, and we will therefore find that receptor methylation constitutes the only noise source on the receptor level. The sampling error quantifies all variability in the output under a constant input, as in \cite{govern_optimal_2014} (see \fref{error}). The final two terms constitute the dynamical error; this is the error that arises from fluctuations in $\ph$ that are caused by differences between past signal trajectories that map onto the same future signal of interest. These fluctuations contribute to the error in $\ph$ because they do not provide any information on the future signal of interest \cite{malaguti_theory_2021} (\fref{error}).

The expression for the prediction error (\eref{prederr}) holds generally for any network in which the signal is inferred from the receptor activity, estimated using a sampling device as in \eref{phat}. Yet, to derive the sensing error $\snrinv$ (see \erefsthree{SNRdef}{errprop}{errdef}) for the chemotaxis network, we need to evaluate the sampling error and the dynamical error, as well as the dynamical gain $\dg$ (see \eref{errprop}). These quantities depend on the specific characteristics of the sensing system and the signal statistics, discussed next.

\begin{figure}[t]
	\centering
	\includegraphics[width=\linewidth]{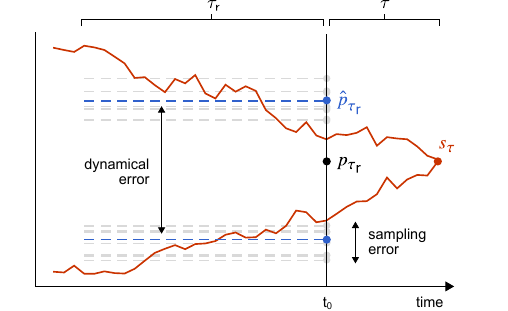}
	\caption{The total error in the cell's estimate of the receptor activity can be decomposed into the dynamical error and the sampling error. For linear signaling systems, a given current or future signal $\st$ (red dot) maps onto a single mean receptor activity $p_{\tr}$ at the current time $t_0$ (black dot) via the dynamic input-output relation of \eref{io} [\fref{intro}(c)]. However, the past input and thus receptor activity on which the estimate $\ph$ (blue dot) is based varies in time, leading to a dynamical error. This error arises because different past trajectories of the signal map onto a common future value $\st$, leading to uninformative variations in $\ph$. Even for a given input trajectory the receptor noise, which in this work is only caused by receptor methylation, and the stochastic nature of the sampling process downstream of the receptor, lead to deviations in the estimate $\ph$ (gray dots) which constitute the sampling error. }
	\flabel{error}
\end{figure}

\subsection{Signal statistics}
\label{sec:sig}
In general, it is hard to know what the natural input statistics are that an organism experiences, and these may vary widely. We can start from the observation that microorganisms in dilute environments are faced with chemical gradients that are exceedingly shallow compared to their own length. In such environments, the only signal property that the cell can measure is the local concentration. But to determine if it is moving in the right direction, the cell must predict the change in concentration over time. So, while the cell can only measure concentrations, it is interested in the concentration's temporal derivative. 

An ideal model system to study networks that can predict temporal changes is the {\it E. coli} chemotaxis network. \textit{E. coli} swims in its environment with a speed which exhibits persistence. This leads to an auto-correlation function for the concentration change  which does not decay instantaneously \cite{mattingly_escherichia_2021}. To model a signal which is characterized by both the concentration and its derivative, and in which correlations in the derivative persist over the correlation time set by the motion of the cell, we use the classical model of a particle in a harmonic well \cite{tjalma2023trade},
\begin{align}
	\delta \dot \ell &= v(t), \elabel{hol}\\
	\dot v &= -\omega_0^2 \delta \ell(t) - v(t)/\tv +\eta_v (t). \elabel{hov}
\end{align}
Here, $\delta \ell(t) \equiv (c(t)-c_0)/c_0$ is the relative deviation of the concentration $c(t)$ from its background value $c_0$. The derivative of this relative concentration is $v(t)$ and $\eta_v(t)$ is a Gaussian white noise. The parameter $\omega_0$ sets the variance in the concentration $\sigma_\ell^2$ relative to that in its derivative $\sigma_\ell^2 = \sigma_v^2/\omega_0^2$, where the variance in the derivative $\sigma_v^2$ is set by the swimming behavior of the cell. The relaxation time $\tv$ is set by the run duration, as this is the timescale over which the input fluctuations decorrelate.

The range of ligand concentrations which \textit{E. coli} might encounter is very large, based on the dissociation constants of the inactive and active receptor conformations. For the Tar-MeAsp receptor ligand combination these respectively are $K_{\text{D}}^{\text{I}}=18\mathrm{\mu  M}$ and $K_{\text{D}}^{\text{A}}=2900 \mathrm{\mu M}$ \cite{sourjik_binding_2002, mello_effects_2007,tu_modeling_2008}. This suggests that the total variance in the ligand concentration is much larger than the concentration change over the course of a run, i.e. $\sigma_\ell \gg \tv\sigma_v$ and thus $\omega_0 \ll \tv^{-1}$. In this regime, the correlation function of $v(t)$ becomes a simple exponential with variance $\sigma_v^2$ and decay time $\tv$: 
\begin{align}
	&\avg{\delta v(t)\delta v(t')} = \sigma_{v}^2 e^{-|t-t'|/\tv}.\elabel{vcor1}
\end{align}

The correlation function of \eref{vcor1} corresponds to what has been observed experimentally for {\it E. coli} cells swimming in shallow exponential concentration gradients \cite{mattingly_escherichia_2021}. When cells swim in shallow gradients, i.e. with a characteristic length much longer than the length of a run, they swim as if there is no gradient. The correlation function of the positional velocity $v_x(t)$ in the absence of a gradient has been measured to be an exponential with variance $\sigma_{v_x}^2$ and decay time $\tv$ set by the duration of a run \cite{mattingly_escherichia_2021}. This can be mapped onto the correlation function of \eref{vcor1}, where $v(t)\equiv c_0^{-1} dc/dt$, when we consider that the concentration gradient is given by $c(t)=c_0 \exp[g x(t)]$ with the gradient steepness $g$. We find for the absolute concentration change over time $dc/dt = dc/dx\,dx/dt = g c(t) v_x(t)$, and thus we have for variance of the relative concentration change $v(t)$:
\begin{align}
	&\sigma_{v}^2 = g^2 \sigma_{v_x}^2. \elabel{vardef}
\end{align}
Experimental measurements provide the relaxation time $\tv^{-1} = 0.86\mathrm{s^{-1}}$ and the variance of the positional derivative $\sigma_{v_x}^2=157 \mathrm{\mu m^2s^{-2}}$ \cite{mattingly_escherichia_2021}.

\subsection{Chemotaxis model}
\label{sec:chem}
 In the {\it E. coli} chemotaxis network, receptors cooperatively control the activity of the kinase CheA, which controls the phosphorylation of the readout protein CheY [\fref{intro}(a)] \cite{maddock_polar_1993,Duke.1999,shimizu_modular_2010,Keegstra.2017}. The receptor cooperativity has been successfully described using the Monod-Wyman-Changeux (MWC) model, where individual receptors are assumed to form clusters in which all receptors must reside in the same activity state \cite{mattingly_escherichia_2021, shimizu_modular_2010, monod_nature_1965,sourjik_functional_2004, mello_allosteric_2005, keymer_chemosensing_2006, tu_modeling_2008, kamino_adaptive_2020}. Furthermore, inactive receptors are methylated by the enzyme CheR, which increases the probability for the cluster to be active, and active receptors are demethylated by CheB. These methylation dynamics ensure that the network exhibits perfect adaptation with respect to the background concentration \cite{segall_temporal_1986, barkai_robustness_1997, yi_robust_2000, mello_perfect_2003, mello_effects_2007, parkinson_signaling_2015}. Therefore, the activity state of the cluster only responds transiently to changes in the input, and reflects the recent change in concentration.

Because both ligand binding and switching between the active and inactive state of the cluster are fast compared to the input, methylation, and phosphorylation dynamics, it is instructive to take a quasi-equilibrium approach and consider the average cluster activity given the methylation level of the cluster and the extracellular ligand concentration. In the linear noise approximation we have for the activity (see appendix \ref{app:chem})
\begin{align}
	a(t) \equiv \avg{n(t)|\delta m,\delta \ell} = p +\alpha \delta m(t) - \beta \delta \ell(t), \elabel{act}
\end{align}
where $p$ is the mean activity, $\delta m(t)$ represents the methylation level of the cluster, and $\delta \ell(t)$ represents the ligand concentration, both defined as deviations from their mean. The constants $\alpha$ and $\beta$ respectively depend on the free energy cost of methylation $\tilde \alpha$, and on the dissociation constants $K_{\text{D}}^{\text{I}}$ and $K_{\text{D}}^{\text{A}}$ and background concentration $c_0$. The methylation dynamics are given by,
\begin{align}
	\dot{\delta m} &= -\delta a(t)/(\alpha \tm) +\eta_{m}(t), \elabel{dmi}
\end{align}
where $\tm$ is the adaptation time, and $\eta_m$ is Gaussian white noise (see \eref{etami}). 

The dynamic gain of the network maps the signal of interest onto the receptor activity [\eref{io}, \fref{intro}(c)]. For the purpose of navigation, we define the signal of interest to be the change in concentration $\vt\equiv v(t_0+\tau)$ some time $\tau\geq0$ into the future. The autocorrelation of the change in concentration is given by \eref{vcor1}. The dynamic gain of the chemotaxis network with respect to this signal of interest is (\erefsrange{appnv}{appdg}),
\begin{align}
	\dg =  \frac{g_{v\to p} e^{-\tau/\tau_v} }{(1+\tm/\tv)(1+\tr/\tv)} =  \frac{-\tm \beta e^{-\tau/\tau_v} }{(1+\tm/\tv)(1+\tr/\tv)}, \elabel{dgchem}
\end{align} 
where $\tv$ is the signal correlation time, $\tr$ is the network response time, $\tm$ is the adaptation time, and the static gain from the input signal derivative $v$ to the steady state activity $p$ is given by
\begin{align}
	g_{v\to p}\equiv \partial_v p =-\tm \beta. \elabel{sg}
\end{align}  
\erefstart{dgchem} shows that the dynamic gain $\dg$ is maximized for a fast response $\tr\ll\tv$, and slow adaptation $\tm\gg \tv$. A longer adaptation time increases the dynamic gain via the static gain (\eref{sg}), because the absolute difference between sequential inputs is on average larger over this longer time. Yet, the dynamic gain saturates as $\tm$ increases:
\begin{align}
	\lim_{\tm \to \infty} \dg = \frac{-\tv \beta  e^{-\tau/\tau_v}}{1+\tr/\tv}. \elabel{dglim}
\end{align}  
In this limit, considering that typically $\tau\leq\tv$ and $\tr \ll \tv$, the dynamic gain is approximately proportional to the signal correlation time $\tv$. The reason is that fluctuations further than $\tv$ in the past cannot affect the mapping from the current signal, which is most correlated to the signal of interest $\vt$, to the current receptor state. Finally, increasing the prediction interval $\tau$ reduces the dynamic gain because the correlation between future signal and sensed input decreases.
 
The cluster covariance under a fixed input signal, the third RHS term in \eref{prederr}, arises from the methylation noise. For the chemotaxis network it is given by (\erefsrange{noisestart}{apprnoise})
\begin{align}
	\E[t_i,t_j,\sp]{\C{n_i(t_i),n_j(t_j)|\sp}} &= \frac{\alpha p(1-p)}{\RT(1+\tr/\tm)}, \nonumber \\
	&\approx\alpha p(1-p)/\RT. \elabel{rnoise}
\end{align}
Here, $p$ is the mean cluster activity and $\RT$ is the total number of independent receptor clusters.  Note that in contrast to previous work \cite{malaguti_theory_2021,malaguti_receptor_2022}, the cluster covariance does not depend on receptor-ligand binding noise because here we have assumed that ligand binding is much faster than the response time $\tr$, setting the receptor-correlation time $\tc$ to zero. Still, the cluster state remains correlated over time due to receptor methylation. Because the methylation - or adaptation - timescale $\tm$ must be longer than the response time $\tr$ for the system to respond to transient changes in the input, the methylation noise cannot be time-averaged like ligand binding noise, i.e. $1+\tr/\tm \approx 1$. Because methylation noise affects the receptor activity via the factor $\alpha$ (\erefstwo{act}{dmi}), the cluster covariance also increases with $\alpha$, as it increases the temporal covariance within each cluster. The only way to mitigate receptor noise is by increasing the number of independent clusters $\RT$.

Finally, the variation in the network output that is caused by the past input trajectory, the term that controls the dynamical error (\eref{prederr}), is given by (\erefsrange{sigstart}{sigfin})
\begin{align}
	&\V{\E[t_i]{\avg{n(t_i)|\sp}}} = \nonumber \\ &\qquad \frac{g_{v\to p}^2 \sigma_v^2}{(1+\tm/\tv)(1+\tr/\tv)}\left(1+\frac{\tm \tr}{\tv(\tm+\tr)}\right), \elabel{sigvar}
\end{align}
with the static gain $g_{v\to p}$ given by \eref{sg}. Just like the dynamic gain (\eref{dgchem}) this variation is maximized for a fast response $\tr \ll \tv$ and slow adaptation $\tm \gg \tv$. Indeed, in the regime that $\tm \gg \tv$ we have $\V{\E[t_i]{\avg{n(t_i)|\sp}}}\propto\tm$. Therefore, unlike the dynamic gain, this variation does not saturate for an increasing adaptation time. The reason is that more and more values of the historical input contribute to the variance in the output as long as the system does not adapt. Not all of the variation quantified by \eref{sigvar} will carry information about the signal of interest $\vt$, and this uninformative variation constitutes the dynamical error (see \eref{prederr}).

Substitution of \erefstwo{rnoise}{sigvar} in \eref{prederr} yields the predicton error, i.e. the error in the estimate of the receptor activity:
\begin{align}
	&\ep = \frac{p^2}{\mN} +\frac{p(1-p)}{\NI} + \dgs \sigma_v^2\bigg[e^{2\tau/\tv}\nonumber \\
	& \times\left(1+\frac{\tm}{\tv}\right)\left(1+\frac{\tr}{\tv}\right)\left(1+\frac{\tm \tr}{\tv(\tm+\tr)}\right) -1\bigg], \elabel{prederrchem}
\end{align} 
with the dynamic gain $\dg$ of \eref{dgchem} and the number of independent samples 
\begin{align}
	\NI\equiv\fI\mN=\frac{\mN}{1+\mN/\RI}, \elabel{ni}
\end{align} 
where $\fI=1/(1+\mN/\RI)$ is the fraction of independent samples and $\RI = \RT(1+\tr/\tm)/\alpha$ is the number of independent receptor states during an integration time $\tr$. The number of independent receptor states decreases with $\alpha$ because it increases the temporal covariance within each cluster (\eref{rnoise}).

\erefstart{ni} reflects that the number of samples $\mN$ and the number of independent receptor states $\RI$ are fundamental resources that limit the accuracy like weak links in a chain \cite{govern_optimal_2014}: when $\mN\gg\RI$ the number of independent samples is limited by the number of receptor states and $\NI\approx\RI$, and vice versa, when $\RI\gg\mN$ the total number of samples is limiting and $\NI\approx\mN$.

\begin{figure*}[t]
	\centering
	\includegraphics[width=\linewidth]{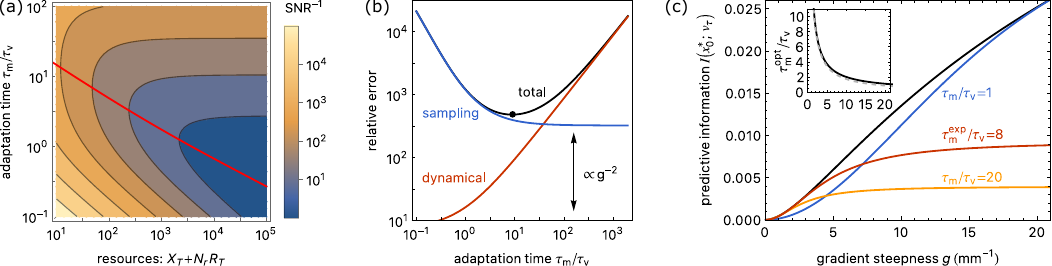}
	\caption{The relative error is set by the resource availability, adaptation time, and gradient steepness. (a) The relative error (\eref{chemerr}) as a function of the resource availability $C=\xT+\Nr \RT$ and the adaptation time $\tm/\tv$. The relative error decreases monotonically with higher resource availability.  The error is minimized for the optimal adaptation times indicated by the red line, which decreases with the resource availability. The ratio of readouts to receptors $\xT/\RT$ obeys \eref{optratchem}. (b) The dynamical error, sampling error, and the total error (\eref{chemerr}), as a function of the adaptation time $\tm/\tv$. The optimal adaptation time arises from a trade-off between the sampling error, which decreases with the adaptation time, and the dynamical error, which increases with the adaptation time. The minimal total error (black dot) occurs close to the point where the sampling saturates as a function of $\tm/\tv$. The minimal sampling error is proportional to $1/g^2$. (c) The predictive information $I(x_0^\ast;\vt)=I(\ph;\vt)=0.5\log(1+\text{SNR})$, with the SNR of \eref{chemerr}, between the current number of phosphorylated readouts $x_0^\ast=\mN \ph$ (\eref{phat}) and the future input derivative $\vt$, for various adaptation times $\tm$. Along the black curve, the adaptation time has been optimized; $\tmopt/\tv$ as a function of the gradient steepness is shown in the inset. Experiments show that for {\it E. coli} the adaptation time is $\tm^\text{exp}/\tv\approx8$ \cite{segall_temporal_1986, shimizu_modular_2010, mattingly_escherichia_2021}, which is close to optimal for $g\lesssim 4 \mathrm{mm^{-1}}$ (red curve). Reducing the adaptation time reduces the accuracy in shallow gradients and increases it in steeper gradients (blue curve), while increasing the adaptation time reduces the accuracy in steeper gradients but does not markedly increase the accuracy in shallow gradients (yellow curve). This suggests that the system has been been optimized for sensing shallow gradients. Inset: the optimal adaptation time $\tmopt/\tv$ scales inversely with the gradient steepness $g$, numerical result (solid black line), and analytical approximation (dashed gray line, \eref{approxtm}). In (a) and (b) $g=2\mathrm{mm^{-1}}$, in (b) and (c) $\xT = 10^4$ and $\RT=8$ \cite{mattingly_escherichia_2021,tjalma2023trade,reinhardt2023path}. Other parameters: $\Nr=12$, $p=0.3$, $f=f^\text{opt}=0.5$, $\tr =0.1\mathrm{s}$, $\tau=\tv=1.16\mathrm{s}$, $c_0 = 100\mathrm{\mu m}$, $\sigma_{v_x}=157\mathrm{\mu m^2 s^{-2}}$ \cite{mattingly_escherichia_2021, tjalma2023trade, shimizu_modular_2010,reinhardt2023path}; $\tilde \alpha=2k_BT$ \cite{shimizu_modular_2010}; $K_{\text{D}}^{\text{I}}=18\mathrm{\mu  M}$ and $K_{\text{D}}^{\text{A}}=2900 \mathrm{\mu M}$ \cite{sourjik_binding_2002, mello_effects_2007,tu_modeling_2008}. Code to reproduce this figure is available publicly \cite{tjalma_zenodo_predicting_2024}.}
	\flabel{results}
\end{figure*}

\subsection{Relative prediction error}
\label{sec:snr}
The central result of this work is the relative error made by the {\it E. coli} chemotaxis network when it predicts the future concentration change. Using the definition of the signal-to-noise ratio (\eref{SNRdef}), with the dynamic gain given in \eref{dgchem}, and the prediction error given by \eref{prederrchem}, we obtain
\begin{widetext}
	\begin{align}
		\snrinv = \underbrace{\frac{e^{2\tau/\tv}}{\tm^2 \beta^2 \sigma_v^2}\left(1+\frac{\tm}{\tv}\right)^2\left(1+\frac{\tr}{\tv}\right)^2\left(\frac{p^2}{\mN} + \frac{p(1-p)}{\NI}\right)}_\text{sampling error} + \underbrace{e^{2\tau/\tv}\left(1+\frac{\tm}{\tv}\right)\left(1+\frac{\tr}{\tv}\right)\left(1+\frac{\tm \tr}{\tv(\tm+\tr)}\right) -1}_\text{dynamical error}. \elabel{chemerr}
	\end{align}
\end{widetext}
This expression is strikingly similar to the relative error of the push-pull network without adaptation, which was derived in earlier work \cite{malaguti_theory_2021}. The reason is that, while the adaptation system affects the receptor dynamics, the downstream push-pull motif still acts as a device that discretely samples the receptor state. As a result, the relative error has two contributions: the sampling error, which arises from the stochasticity in sampling the state of the receptor, and the dynamical error, which arises from the dynamics of the input signal (see \fref{error}). However, while this expression for the relative error has a form that is similar to that for the push-pull network, there are also key differences.
	
First of all, both the sampling and the dynamical error depend on the forecast interval. In general, the dynamical error arises because while the system aims to predict the current or future derivative, it measures the change in concentration over the timescale $\tm$ on the level of the receptor, and reads out the receptor activity over the timescale $\tr$ (\fref{error}). The network thus only measures an instantaneous concentration change when both $\tm$ and $\tr$ go to zero. Still, even in this limit, the dynamical error remains finite as long as the forecast interval is larger than zero, due to the inherent unpredictability of the future signal. 

Perhaps surprisingly, the relative sampling error also depends on the forecast interval $\tau$. While the absolute sampling error of the network is independent of the forecast interval (\eref{prederrchem}), the dynamic gain does depend on it (\eref{dgchem}). When the forecast interval increases, the dynamic gain decreases, reducing the effect of the signal of interest on the receptor activity. Therefore, while the absolute sampling error remains constant, the relative sampling error increases with the forecast interval. In short, for a larger forecast interval it becomes harder to lift the signal above the sampling noise.

The second notable difference with the result on the push-pull network concerns the role of adaptation. It reflects the fact that the chemotaxis system takes a temporal derivative at the receptor level on a timescale set by the adaptation time. The dynamical error increases monotonically with the adaptation time $\tm$, because for a longer adaptation time the system compares the current concentration to concentrations further in the past. Consequently, this change in concentration is less informative about the current derivative, which is the signal property most correlated to the future derivative (\eref{vcor1}). In contrast, the sampling error decreases monotonically with $\tm$, because a longer adaptation time increases the dynamic gain (\eref{dgchem}). How the optimal adaptation time that arises from these antagonistic effects depends on other parameters, such as the gradient steepness and the resource availability, is discussed in section \ref{sec:opttm}.

A third difference resides in the number of independent samples $\NI$ (\eref{ni}). For a push-pull network driven by a simple receptor, the number of independent samples is given by $\NIppn=\fI^\text{PPN}\Neff$, where the number of effective samples $\Neff=\mN$ in the irreversible limit, as we also study here \cite{govern_optimal_2014, malaguti_theory_2021,malaguti_receptor_2022}. For the push-pull network the fraction of independent samples can be expressed as
$\fI^\text{PPN} = 1/(1+\mN/\RI^\text{PPN})$
with the number of independent receptor states during an integration time $\RI^\text{PPN} = \RT(1+\tr/\tc)$, where $\tc$ is the correlation time of the receptor binding state \cite{govern_optimal_2014, malaguti_theory_2021,malaguti_receptor_2022}. In our treatment of the chemotaxis model we consider the limit where $\tc \ll \tr$, in which case $\RI^\text{PPN}$ diverges and $\fI^\text{PPN}\approx 1$. However, the fraction of independent samples does not become unity for the chemotaxis network because the receptor state remains correlated due to the slow methylation dynamics, $\tm \gg \tr$. Therefore, the number of independent receptor states becomes limited by the number of receptor clusters and their covariance $\RI\approx\RT/\alpha$ (\eref{ni}).

The sampling error can be mitigated in a number of ways. One is to increase the number of receptors per cluster $\Nr$, because this increases the magnitude of the static gain (see \eref{sg} where $\beta\propto\Nr$). Another is to simultaneously increase the total number of samples $\mN$ and the number of independent samples $\NI$,  which requires increasing both the number of readout molecules $\xT$ and the number of receptor clusters $\RT$ (\erefstwo{mn}{ni}). However, increasing the cluster size, the number of clusters, or the number of readout molecules, all require a larger number of proteins to be used by the network, which are resources that come at a physical cost.

\subsection{Optimal resource allocation}
 To investigate how resources should be optimally allocated to minimize the sampling error (\eref{chemerr}) we define a simple cost function, as in \cite{tjalma2023trade}:
\begin{align}
	C=\xT+\Nr \RT,
\end{align}
where $\xT$ is the number of readout molecules, $\RT$ is the number of independent receptor clusters, and $\Nr$ is the number of receptors per cluster. This cost function captures the idea that a cell must choose whether it spends its resources on making more readout molecules on the one hand, or more receptors on the other. Running the network also requires chemical power to drive the methylation and phosphorylation cycles. However, earlier work has shown that including the cost of driving the network does not significantly alter the optimal design of the network \cite{tjalma2023trade}. Here, we therefore omit these running costs.

Given a total resource availability $C$ and a fixed number of receptors per cluster, the cell can tune the ratio of receptors to readouts. To determine what the optimal ratio is that minimizes the sampling error, we express both $\RT$ and $\xT$ in terms of their ratio and the total resource availability $C$, and we use that we can express the mean number of samples as in \eref{mn}. Subsequently taking the derivative of \eref{chemerr} with respect to $\xT/\RT$ and equating to zero then gives the optimal ratio,
\begin{align}
	\left(\frac{\xT}{\RT}\right)^\text{opt} =\frac{\sigma_X}{\sigma_R} \frac{p \sqrt{1+\tr/\tm}}{f(1-f)} \approx \frac{\sigma_X}{\sigma_R} \frac{p }{f(1-f)}, \elabel{optratchem}
\end{align}
where we have used that the adaptation time must be larger than the response time and thus $\sqrt{1+\tr/\tm}\approx 1$. We have further defined the noise per receptor $\sigma_R^2\equiv \alpha p(1-p)/\Nr = \tilde{\alpha}p^2(1-p)^2$ (also see \erefsthree{appdfm}{appa}{apprnoise}), and the noise per readout molecule $\sigma_X^2\equiv f(1-f)$. In terms of $\mN$, using \eref{mn}, we find that \eref{optratchem} yields an intuitive relation for optimal networks,
\begin{align}
	\mN = \frac{\sigma_X}{\sigma_R}\RT. \elabel{optrel}
\end{align}
This relation shows that for equal noise magnitudes per protein, the average number of samples should equal the total number of receptor clusters. This simple relation arises from the fact that the methylation noise cannot be averaged out, and a minimally redundant design is therefore one in which each receptor cluster is sampled once.

Given the optimal ratio of readouts to receptors in \eref{optratchem} we can compute the relative error (\eref{chemerr}) as a function of the total resource availability $C$ and the adaptation time $\tm$ [\fref{results}(a)]. As expected, we find that the error decreases monotonically with the resource availability. More interesting is that we find a clear optimum for the adaptation time $\tm$.

\subsection{Optimal adaptation time}
\label{sec:opttm}
The optimal adaptation time $\tm$, given by the red line in \fref{results}(a), arises from the antagonistic effect of the adaptation time on the sampling error and the dynamical error [\fref{results}(b)]. The sampling error decreases monotonically with the adaptation time because a longer adaptation time increases the change in the receptor activity upon the same change in the current or future signal derivative, i.e. it increases the (dynamic) gain (\erefsthree{dgchem}{sg}{chemerr}). However, increasing the adaptation time means that the derivative is taken over a longer time further back into the past, and this derivative will be less informative about the future derivative that the cell aims to predict: the dynamical error increases monotonically with $\tm$ (\eref{chemerr}). The minimal total error occurs for the smallest adaptation time that is sufficiently large to lift the signal above the noise, i.e. reduce the sampling error, while minimizing the dynamical error [\fref{results}(b)].

The value of the adaptation time for which the total error is minimized depends on the resource availability $C$ and the gradient steepness $g$: these parameters set the magnitude of the sampling error [\eref{chemerr} and \fref{results}(b)]. To obtain analytical insight into the optimal adaptation time $\tmopt$, we exploit that the response time $\tr$ must be smaller than the adaptation time $\tm$ to mount a non-zero response to transient input changes. We further consider that the relevant regime for {\it E. coli} is likely that where gradients are shallow relative to the length of a run (also see section \ref{sec:exp}). This means that the sampling error dominates over the dynamical error (see \eref{chemerr} where $\sigma_v^2=g^2\sigma_{v_x}^2$, \eref{vardef}). In this regime the adaptation time must be large relative to the signal correlation time $\tv$, which is set by the duration of a run. We obtain for the optimal adaptation time (see appendix \ref{app:opttm})
\begin{align}
	\tmopt \approx \frac{\sqrt{2}}{\beta g \sigma_{v_x}}\sqrt{\frac{p^2}{\mN}+\frac{p(1-p)}{\NI}}, \quad \text{for }\tm \gg \tv, \tr, \elabel{approxtm}
\end{align}
where the number of independent samples $\NI$ is given by \eref{ni} with $\RI=\RT/\alpha$. The inset of \fref{results}(c) shows that \eref{approxtm} is a good approximation of the optimal adaptation time over a large range of the gradient steepness $g$.

\subsection{Comparison to experiment}
\label{sec:exp}
To check whether the uncovered design principles (\erefstwo{optratchem}{approxtm}) are relevant to real world biochemical networks, we evaluate the design of the {\it E. coli} chemotaxis network in this light. 

To assess the design principle of \eref{optratchem}, we use the definitions of $\sigma_X$ and $\sigma_R$ given below it. For $p$ and $f$ of order $1/2$ and $\tilde{\alpha} = 2$, based on experiment \cite{shimizu_modular_2010}, \eref{optratchem} predicts an optimal number of readout molecules per receptor cluster of $\xT/\RT\approx 3$. This is in good agreement with earlier predictions \cite{govern_optimal_2014} and the experimental data of Li and Hazelbauer \cite{li_cellular_2004}, assuming a cluster consists of $2$ trimers of receptor dimers and $2$ CheA dimers \cite{briegel2012bacterial}. With $\xT\sim 10^3 - 10^4$ readout molecules depending on the growth rate \cite{li_cellular_2004}, this result, i.e. $\xT/\RT\approx 3$, suggests that the number of receptor clusters is in the range $\RT \sim 10^2 - 10^3$. On the other hand, fitting more recent experimental data with an MWC based chemotaxis model as we use here, suggests a much smaller number of receptor clusters of $\RT\approx 8$ \cite{mattingly_escherichia_2021, tjalma2023trade, reinhardt2023path}. However, this estimate for the number of receptor clusters was based on fitting the  noise amplitude of the model \cite{tjalma2023trade,reinhardt2023path} to the experimental data of \cite{mattingly_escherichia_2021}. Recent experiments indicate that the receptor array is poised to a critical point \cite{keegstra_near_2023}, where receptor switching becomes correlated over long distances, and it is conceivable that this small value of $\RT\approx 8$ corresponds to the small number of domains over which the receptors effectively switch in concert. More work is needed to understand whether receptor switching near a critical point can effectively be described by an MWC model, and whether the design rule unveiled here (\eref{optratchem}), also generalizes to a receptor array near a critical point. Lastly, further study is necessary to understand whether information transmission in this system is maximized near a critical point \cite{meijers2021behavior}.

The adaptation time of the {\it E. coli} chemotaxis system has repeatedly been shown to be $\sim10\mathrm{s}$, yielding $\tm^\text{exp}/\tv \approx 8$ \cite{segall_temporal_1986, shimizu_modular_2010, mattingly_escherichia_2021}. Given the estimated resource allocation in the effective MWC description, $\xT=10^4$ and $\RT=8$ with $\Nr=12$ \cite{mattingly_escherichia_2021,tjalma2023trade,reinhardt2023path}, this adaptation time is close to optimal for gradient steepnesses  $g \lesssim 4\mathrm{mm^{-1}}$ [\fref{results}(c)]. In particular, while decreasing the methylation time improves the prediction accuracy in steeper gradients, it reduces information transmission in shallower gradients. On the other hand, while increasing the methylation time beyond the measured one decreases the accuracy in steeper gradients, the improvement in shallow gradients is only very minor because the system is already very close to the fundamental bound on the predictive information as set by the resource constraint and the gradient steepness [\fref{results}(c)]. These arguments show that the methylation time of {\it E. coli} is indeed optimal for sensing shallow gradients with $g \lesssim 4\mathrm{mm^{-1}}$. It suggests that the chemotaxis system has been optimized for navigating weak gradients. To get an idea of what this gradient steepness means we can compare it to the length of an {\it E. coli} cell, which is $\sim 1\mathrm{\mu m}$. To cover a gradient length scale $g^{-1}=1/4\mathrm{mm}$ thus requires the cell to move at least $250$ times its body length, corresponding to approximately $10$ runs in the same direction \cite{staropoli2000computerized, turner2000real, darnton2007torque}. This illustrates how extremely shallow the gradients that {\it E. coli} can encounter likely are. Moreover, it suggests that it is most important to maximize accuracy in shallow gradients, where it is hard to distinguish signal from noise. In steeper gradients {\it E. coli} would be further from the optimal design, but the total information it obtains about the signal of interest is still larger because the input fluctuations are bigger.

\section{Discussion}
Microorganisms that navigate chemical gradients need to predict the concentration change that they will encounter. For simple input signals where the change in concentration is Markovian, the optimal way to achieve this is to measure the current time derivative of the concentration \cite{tjalma2023trade}. Measuring such temporal concentration changes requires perfect adaptation. Moreover, to measure the most recent concentration change, the adaptation time must be short relative to the correlation time of the input. However, building and maintaining a biochemical network costs physical resources. When the resource availability is limited, the signal is obscured by noise in the network. The only way to lift the signal above the noise in this regime, is to increase the adaptation time. This trade-off between lifting te signal above the noise, and measuring a concentration change which is informative of the future input, sets the optimal adaptation time. 

The optimal adaptation time depends on the amount of resources available to maintain the network, and the magnitude of changes in the input. The latter is set by the swimming behavior of the cell and the steepness of the chemical gradient it navigates. In steeper gradients the input changes more strongly, which reduces the sampling error and increases the signal-to-noise ratio. A smaller sampling error allows for a shorter adaptation time, which mitigates the dynamical error and maximizes the overall accuracy. Therefore, the optimal adaptation time to predict the concentration change scales approximately inversely with the gradient steepness. Interestingly, simulations show that the optimal adaptation time that maximizes navigational performance also increases as the gradient becomes more shallow \cite{frankel2014adaptability, dufour2014limits}. This indicates that predicting the concentration change is indeed important for successful navigation, in line with results of agent-based simulations on the interplay between prediction and navigation \cite{becker_optimal_2015}. 
	
Our analysis provides a possible explanation for a puzzling observation. During chemotaxis, {\it E. coli} performs subsequent runs of approximately one second in different directions. Runs in the correct direction relative to the gradient are extended, and vice versa, such that the cell moves up a gradient of attractant on average. To implement this strategy, {\it E. coli} must predict how the concentration will change while it navigates the gradient. To this end, it seems natural to measure the change in concentration over the course of one run, i.e. over approximately one second. However, the adaptation time of  {\it E. coli} is around ten seconds \cite{segall_temporal_1986, shimizu_modular_2010, mattingly_escherichia_2021}. This raises the question, why would {\it E. coli} measure concentration changes over a timescale that is much longer than that of a run? Our work shows that the adaptation time must be this long to discern the signal from the inevitable biochemical signaling noise in shallow gradients.

More generally, our results provide insight into the optimal design of adaptive signaling networks. First and foremost, this improves our understanding of navigation behavior of microorganisms. But the uncovered principles might well hold more generally and shed light on other adaptive signaling networks as well, e.g. that of rod cells in the vertebrate eye \cite{alberts2008cell}. Moreover, our theory facilitates the optimal design of micro-robots that need to navigate environments without a map.

\begin{acknowledgments}
We thank Vahe Galstyan for a careful reading of the manuscript. This work is part of the
Dutch Research Council (NWO) and was performed at
the research institute AMOLF. This project has received
funding from the European Research Council (ERC)
under the European Union's Horizon 2020 research
and innovation program (grant agreement No. 885065).
\end{acknowledgments}

\appendix

\section{The prediction error}
\label{app:prederr}
Here we derive the general expression for the prediction error $\ep$, which shows how the complete error decomposes into independent parts caused by fluctuations in the number of samples $N$, the error of a sampling process with a fixed number of samples and a constant input, and uninformative fluctuations from the input signal. Our starting point is the decomposition of the error in \eref{errgen}.

The first term of \eref{errgen} is straightforward to compute, using the definition of $\ph$ from \eref{phat} we obtain,
\begin{align}
	\V{\E{\ph|N}} &= \V{\E[t_i,n_i]{\frac{1}{\mN}\sum_{i=1}^N n_i(t_i)|N}}_N, \elabel{apperror11}\\
	&= \frac{1}{\mN^2}\V{N \E[t_i]{\avg{n(t_i)}}}_N, \elabel{apperror12}\\
	&= \frac{p^2}{\mN},\elabel{apperror13}
\end{align}
where the subscripts after the expected values and variances denote the random variables over which the expectation is taken. For instance, in \eref{apperror11} the expected value is taken under a fixed number of samples $N$ over the state $n_i \in \{0,1\}$ of each cluster, later also denoted with angle brackets as an ensemble average, and over alle sampling times $t_i$, which are exponentially distributed with PDF \cite{govern_optimal_2014}
\begin{align}
	f(t_i)=\frac{1}{\tr}e^{-(t_0-t_i)/\tr}. \elabel{tidist}
\end{align}
From \eref{apperror12} to \eref{apperror13} we use that the average number of active receptor clusters is defined as $p\equiv \E[t_i]{\avg{n(t_i)}}$, which is constant with respect to $N$. The variance is subsequently taken over the Poisson distributed number of samples $N$, with both mean and variance $\mN$. The resulting expression (\eref{apperror13}) is the error that arises because the network cannot distinguish between those readout molecules that sampled an inactive cluster, and those that did not sample a cluster at all \cite{malaguti_receptor_2022,govern_optimal_2014}.

We decompose the second term of \eref{errgen} in two steps. First, we use the definition of $\ph$ (\eref{phat}) and split the self- and cross-terms in the covariance of the kinase activity:
\begin{align}
	&\E{\V{\ph|N}} = \E[N]{\V{\frac{1}{\mN}\sum_{i=1}^N n_i(t_i)|N}_{n_i,t_i}}, \elabel{Nstep}\\
	&= \frac{1}{\mN^2}\mathbb{E}[N\V{n_i(t_i)} \nonumber \\
	& \qquad+ N(N-1)\C{n_i(t_i),n_j(t_j)}]_N, \elabel{Nstep1}\\ 
	& = \frac{p(1-p)}{\mN} +  \C{n_i(t_i),n_j(t_j)}. \elabel{Nstep2}
\end{align}
From \eref{Nstep1} to \eref{Nstep2} we used that both the variance of each cluster and the covariance between clusters are independent of the number of samples $N$, and that for a Poisson distributed number of samples $N$ we have $\E{N(N-1)}=\mN^2$. To continue, the covariance between different kinases at different times can be decomposed into contributions from the receptor noise, and fluctuations in the full history of the input signal, the trajectory $\sp$,
\begin{align}
	&\C{n_i(t_i)n_j(t_j)} = \E[t_i,t_j,\sp]{\C{n_i(t_i),n_j(t_j)|\sp}} \nonumber \\
	&\qquad + \C{\E[t_i]{\avg{n_i(t_i)|\sp}},\E[t_j]{\avg{n_j(t_j)|\sp}}}_{\sp}, \elabel{covdec0} \\
	&= \E[t_i,t_j,\sp]{\C{n_i(t_i),n_j(t_j)|\sp}} + \V{\E[t_i]{\avg{n(t_i) |\sp}}}_{\sp}, \elabel{covdec}
\end{align}
where we use that $\E[t_i]{\avg{n_i(t_i)|\sp}}=\E[t_j]{\avg{n_j(t_j)|\sp}}$. The two terms on the RHS of \eref{covdec} respectively describe the covariance between clusters when the input is fixed, and the variance that is caused by input fluctuations. The first term is the receptor-level noise, which for the chemotaxis model considered in this work arises only from methylation \erefsrange{noisestart}{apprnoise}. The second term is the variance of the mean activity conditional on the input, which is the signal-induced variance. This signal induced variance comprises all variance caused by the input, so both the dynamical error and the variance that is informative of the signal of interest $\dg^2\sst$ \erefsrange{sigstart}{sigfin}.

Combining \erefstwo{Nstep2}{covdec} gives
\begin{align}
	\E{\V{\ph|N}} =& \frac{p(1-p)}{\mN} \nonumber \\
	&+ \E[t_i,t_j,\sp]{\C{n_i(t_i),n_j(t_j)|\sp}} \nonumber \\
	&+ \V{\E[t_i]{\avg{n(t_i)|\sp}}}_{\sp}. \elabel{error2}
\end{align}

Finally, the third term of \eref{errgen} is the contribution of the signal of interest to the output variance:
\begin{align}
	\V{\E{\ph|\st}} &= \V{\E[t_i,n_i,N]{\frac{1}{\mN}\sum_{i=1}^N n_i(t_i)|\st}}_{\st}, \\
	&= \V{\E[t_i]{\avg{n(t_i)|\st}}}_{\st}, \\
	&=\dg^2\sst, \elabel{error3}
\end{align}
where in the last step we have used the dynamic input output relation of \eref{io}. The dynamic gain $\dg$ of the chemotaxis network is derived in \erefsrange{appnv}{appdg}. Substituting the equalities of \erefsthree{apperror13}{error2}{error3} in \eref{errgen} of the main text gives the complete prediction error given in \eref{prederr} in the main text.

We note that this derivation deviates from that of Malaguti and Ten Wolde \cite{malaguti_receptor_2022} in that \eref{Nstep} includes the contributions from all signal variations, including the informative signal variations  (which are then subtracted from the full variance in \eref{errgen}), while in  \cite{malaguti_receptor_2022} the corresponding term does not contain these informative signal fluctuations. While the final result is identical, the derivation presented here is arguably easier.

\section{The chemotaxis network}
\label{app:chem}
In the \textit{E. coli} chemotaxis network, receptors cooperatively control the activity of the kinase CheA, and the activity is adaptive due to the methylation of inactive receptors \cite{maddock_polar_1993,Duke.1999,shimizu_modular_2010,Keegstra.2017, segall_temporal_1986,parkinson_signaling_2015}. We here follow the widely used approach to describe the effects of receptor cooperativity and methylation on kinase activity via the Monod-Wyman-Changeux (MWC) model \cite{monod_nature_1965,sourjik_functional_2004,mello_allosteric_2005,keymer_chemosensing_2006,mello_effects_2007,tu_modeling_2008,shimizu_modular_2010,kamino_adaptive_2020,mattingly_escherichia_2021}. In this model, each receptor can switch between an active and inactive conformational state $n$ and receptors are partitioned into clusters of equal size $\Nr$. In the spirit of the MWC model, receptors within a cluster switch conformation in concert, so that each cluster is either active or inactive  \cite{monod_nature_1965}. Furthermore, it is assumed that receptor-ligand binding and conformational switching are faster than the other timescales in the system, such that the activity state of the receptor can effectively be described by its equilibrium probability to be active, given the methylation level of the cluster $m$ and the external ligand concentration $\ell$. The probability for the receptor cluster to be active is then described by:
\begin{align}
	a(\ell, m)\equiv \avg{n|\ell, m} = (1+\exp(\Delta F_T(\ell , m)))^{-1},
\end{align}
where $\Delta F_T(\ell, m) = -\Delta E_0 + \Nr(\Delta F_\ell(\ell) + \Delta F_m(m))$ is the free-energy difference between the active and inactive state, which is a function of free-energy difference arising from ligand binding and methylation:
\begin{align}
	&\Delta F_\ell(\ell) = \ln(1+\ell(t)/\KDI) -\ln(1+\ell(t)/\KDA),\\
	&\Delta F_m(m) = \tilde \alpha(\bar m -m(t)). \elabel{appdfm}
\end{align}
Between the two states the cluster has an altered dissociation constant, which is denoted $\KDI$ for the inactive state, and $\KDA$ for the active state. The free-energy difference due to methylation has been experimentally shown to depend approximately linearly on the methylation level \cite{shimizu_modular_2010}. We assume that inactive receptors are irreversibly methylated, and active receptors irreversibly demethylated, with zero-order ultrasensitive kinetics \cite{tu_modeling_2008,Emonet.2008,tostevin_mutual_2009}. The methylation dynamics of a receptor cluster is then given by:
\begin{equation}
	\begin{split}
		\dot m =& (1-a(\ell, m)) k_R - a(\ell, m) k_B + B_{m}(a)\xi(t),
	\end{split}
\end{equation}
with $B_m(a) = \sqrt{(1-a(\ell, m)) k_R + a(\ell, m) k_B}$, and unit white noise $\xi(t)$. These dynamics indeed give rise to perfect adaptation, since from this equation we find that the steady state cluster activity is given by $p\equiv\bar a = 1/(1+k_B/k_R)$, thus indeed independent of the ligand concentration. 

In this work we consider linear dynamics, we therefore employ a linear noise approximation \cite{vanKampen1992}. The deviation of the equilibrium cluster activity from its mean $\delta a(t) = a(t) - p$ is then given by
\begin{align}
	\delta a(t) \equiv \avg{n(t)|\delta\ell,\delta m}-p= \alpha \delta m(t) - \beta \delta \ell(t), \elabel{appa}
\end{align}
with $\alpha = \tilde \alpha \Nr p(1-p)$ and $\beta =\kappa \Nr p(1-p)$, with $\kappa= (1 +\KDI/c_0)^{-1}-(1 +\KDA/c_0)^{-1}$. For the methylation dynamics on one cluster we then obtain,
\begin{align}
	\dot{\delta m} &= -\delta a(t)/(\alpha \tm) +\eta_{m}(t), \elabel{appm}
\end{align}
where we have introduced the adaptation time $\tm=(\alpha(k_R+k_B))^{-1}$ and $\eta_{m}(t)$ is Gaussian white noise on a single cluster with correlation function 
\begin{align}
	\avg{\eta_{m_i}(t)\eta_{m_j}(t')}=\delta_{ij}\delta(t-t') \frac{2 p(1-p)}{\alpha \tm} \elabel{etami}
\end{align} 
between the $i^{th}$ and $j^{th}$ receptor cluster, where $\delta_{ij}$ is the Kronecker delta.
Combining \erefstwo{appa}{appm} yields the change in activity over time
\begin{align}
	\dot{\delta a} = -\delta a(t)/\tm -\beta v(t) +\alpha \eta_m(t),\elabel{appa2}
\end{align}
where we have the change in concentration over time $v(t)\equiv\dot{\delta\ell}$. Using \eref{appa2} we can also express the instantaneous activity as
\begin{align}
	\delta a(t) = \int_{-\infty}^t dt' \left(\alpha \eta_m(t')-\beta v(t')\right)e^{-(t-t')/\tm}. \elabel{appa3}
\end{align}
This expression shows that the cluster activity, when we average out the methylation noise, reflects the change in concentration weighted exponentially over the past adaptation time $\tm$.

\subsection*{Dynamic gain}
The dynamic gain of the network can be obtained by deriving the average response of the network to the signal of interest $\st$. In general we have the expression given in \eref{io} for the dynamic input output relation of linear signaling networks. In our case the signal of interest is the future concentration derivative $\st=\vt$. Using \erefstwo{appa}{appa3}, we find for the average conditional activity,
\begin{align}
	&\avg{n(t_i)|\vt} = \E[\delta\ell,\delta m]{\avg{n(t_i)|\vt,\delta\ell,\delta m}}, \elabel{appnv}\\
	&=p -\beta \int_{-\infty}^{t_i} dt \avg{v(t)|\vt}e^{-(t_i-t)/\tm}, \\
	&= p-e^{-(t_0+\tau-t_i)/\tv}\frac{\tm \beta \vt}{1+\tm/\tv}
\end{align}
where we used that the conditional mean derivative is $\avg{v(t)|\vt}=\vt \exp(-(t_0+\tau-t)/\tv)$, also see \eref{vcor1}. Averaging over all sampling times, distributed as in \eref{tidist}, gives
\begin{align}
		\E[t_i]{\avg{n(t_i)|\vt}} =p- \frac{\tm \beta e^{-\tau/\tv} \vt}{(1+\tm/\tv)(1+\tr/\tv)}. \elabel{appdg}
\end{align} 
Comparison to \eref{io} yields the dynamic gain $\dg$ given in \eref{dgchem}.

\subsection*{Receptor noise}
The variance that is caused by receptor-level (here methylation) noise is the covariance between clusters under a fixed input trajectory, i.e. the first term of \eref{covdec}. We can write this covariance in terms of the equilibrium activity as follows, using \eref{appa} and noting that $\delta\ell(t)$ is contained in $\sp$  for $t\leq t_0$:
\begin{align}
 	&\E[t_i,t_j,\sp]{\C{n_i(t_i),n_j(t_j)|\sp}} \nonumber \\
 	&= \E[t_i,t_j,\sp,\delta m]{\avg{n_i(t_i)n_j(t_j)|\sp, \delta m}} - p^2, \elabel{noisestart}\\ 
 	&= \E[t_i,t_j,\sp,\delta m]{\avg{n_i(t_i)|\sp, \delta m}\avg{n_j(t_j)|\sp, \delta m} - p^2}, \elabel{appnoise2}\\ 
 	&= \E[t_i,t_j,\sp,\delta m]{\avg{\delta a_i(t_i)\delta a_j(t_j)|\sp}}. \elabel{appnoise3}
\end{align}
In \eref{noisestart} we condition on- and average over $\delta m$ to make the connection between the instantaneous cluster state $n_i$ and the cluster activity $a_i$ (\eref{appa}). Then in \eref{appnoise2} we use the fact that when conditioned on both the signal and the methylation level, the cluster states are independent. The covariance in the cluster activity conditioned on the full past input trajectory (\eref{appnoise3}) depends only on the methylation noise, using \erefstwo{appa3}{etami} and keeping the sampling times fixed,
\begin{align}
	&\E[\sp, \delta m]{\avg{\delta a_i(t_i)\delta a_j(t_j)|\sp}} =\mathbb{E}\bigg[\alpha^2 \int_{-\infty}^{t_i} dt \int_{-\infty}^{t_j} dt' \nonumber \\ 
	&\quad \avg{\eta_{m_i}(t)\eta_{m_j}(t')}e^{-(t_i-t)/\tm} e^{-(t_j-t')/\tm}\bigg]_{\sp, \delta m} \\
	&= \avg{\delta_{ij}}\frac{2 \alpha p(1-p)}{\tm} \int^{t^-}_{-\infty}dt e^{-(t^--t)/\tm} e^{-(t^+-t)/\tm}, \\
	&= \frac{\alpha p(1-p)}{\RT}e^{-|t_i-t_j|/\tm},\elabel{apprnoise}
\end{align}
where $t^+ \equiv \max(t_i,t_j)$ and $t^-\equiv \min(t_i,t_j)$, and the number of receptor clusters $\RT$ arises as the average Kronecker delta over all clusters: $\avg{\delta_{ij}}=1/\RT$. Averaging over the exponentially distributed sampling times $t_i$ and $t_j$ (both following \eref{tidist}), yields the receptor noise given in \eref{rnoise}.

\subsection*{Signal induced correlations}
The covariance in the output caused by the variation in the past input signal is given by the second term of \eref{covdec}. It describes all variance in the output caused by input fluctuations, so it comprises both the dynamical error and the informative part $\dgs \sst$. We rewrite the instantaneous activity to the equilibrium activity using \eref{appa} and considering that $\delta\ell(t)$ is contained in $\sp$  for $t\leq t_0$:
\begin{align}
	 &\V{\E[t_i]{\avg{n(t_i) |\sp}}}_{\sp} = \V{\E[t_i, \delta _m]{\avg{n(t_i) |\sp,\delta m}}}_{\sp},  \elabel{sigstart}\\
	 &= \V{p+\E[t_i, \delta _m]{\avg{\delta a(t_i) |\sp}}}_{\sp}, \\
	 &=\V{ -\frac{\beta}{\tr} \int_{-\infty}^{t_0}dt_i\int_{-\infty}^{t_i} dt v(t) e^{-(t_i-t)/\tm} e^{-(t_0-t_i)/\tr}} \elabel{varcondmean}
\end{align}
where in \eref{sigstart} we again condition on- and average over $\delta m$ to make the connection between $n(t)$ and $a(t)$ (\eref{appa}). In \eref{varcondmean} we used \eref{appa3} and the sampling time distribution of \eref{tidist}. Using the correlation function of the concentration derivative, \eref{vcor1}, we continue from \eref{varcondmean} to obtain
\begin{align}
	&\V{\E[t_i]{\avg{n(t_i) |\sp}}}_{\sp} = \frac{\sigma_v^2\beta^2}{\tr^2}\int_{-\infty}^{t_0}dt_i\int_{-\infty}^{t_0}dt_j \bigg( \nonumber \\
	&\int_{-\infty}^{t_i} dt \int_{-\infty}^{t_j} dt' e^{-|t-t'|/\tv} e^{-(t_i-t)/\tm}e^{-(t_j-t')/\tm} \nonumber \\
	&\quad e^{-(t_0-t_i)/\tr} e^{-(t_0-t_j)/\tr}\bigg).
\end{align}
First we perform the integrals over $t$ and $t'$, which yields,
\begin{align}
	&\V{\E[t_i]{\avg{n(t_i) |\sp}}}_{\sp}=\frac{\sigma_v^2\beta^2/\tr^2}{1/\tv^2-1/\tm^2}\int_{-\infty}^{t_0}dt_i\int_{-\infty}^{t_0}dt_j  \nonumber \\
	&\left( \frac{\tm}{\tv} e^{-|t_i-t_j|/\tm}- e^{-|t_i-t_j|/\tv}\right) e^{-(t_0-t_i)/\tr} e^{-(t_0-t_j)/\tr}. 
\end{align}
	Finally, computing the integrals over the sampling times $t_i$ and $t_j$ gives,
\begin{align}
	&\V{\E[t_i]{\avg{n(t_i) |\sp}}}_{\sp} \nonumber \\ 
	&\quad =\frac{\tm^2 \beta^2  \sigma_v^2(1+\tr/\tm+\tr/\tv)}{(1+\tm/\tv)(1+\tr/\tv)(1+\tr/\tm)},  \elabel{sigfin}
\end{align}
which is equivalent to the expression in main text \eref{sigvar} with the static gain of \eref{sg}.
	
\section{Optimal adaptation time}
\label{app:opttm}
Here we give a comprehensive derivation of the approximate optimal adaptation time. To gain analytical insight into the optimal adaptation time we first consider that the adaptation time $\tm$ must be larger than the response time $\tr$ to yield a non-zero response to transient input changes. Subsequently taking the derivative of \eref{chemerr} with respect to $\tm$ then gives,
\begin{align}
	&\frac{\partial \snrinv}{\partial \tm} =\frac{e^{2\tau/\tv}}{\tv}\left(1+\frac{\tr}{\tv}\right)^2\bigg[1- \nonumber\\ 
	&\quad \frac{2\left(1+\tv/\tm\right)}{(\tm \beta g \sigma_{v_x})^2}\left(\frac{p^2}{\mN}+\frac{p(1-p)}{\NI}\right)\bigg], \text{ for }\tm\gg\tr, \elabel{appdsnr}
\end{align}
where the number of independent samples $\NI$ is given by \eref{ni} with $\RI=\RT/\alpha$. Now considering that for {\it E. coli} the adaptation time is much larger than the signal correlation time gives, up to the prefactor,
\begin{align}
	&\frac{\partial \snrinv}{\partial \tm} \propto 1-\frac{2}{(\tm \beta g \sigma_{v_x})^2}\left(\frac{p^2}{\mN}+\frac{p(1-p)}{\NI}\right), \elabel{appdsnr2}
\end{align}
for $\tm\gg\tr,\tv$. Equating \eref{appdsnr2} to zero and solving for $\tmopt$ yields one positive solution, given in \eref{approxtm}.

\bibliography{library_age}

\end{document}